# Near real-time enumeration of live and dead bacteria using a fibre-based spectroscopic device

Authors: Fang Ou[1,2], Cushla McGoverin[1,2], Simon Swift[3], Frédérique Vanholsbeeck[1,2]

[1] Department of Physics, The University of Auckland, Auckland, New Zealand

[2] The Dodd-Walls Centre for Photonic and Quantum Technologies, New Zealand

[3] School of Medical Sciences, The University of Auckland, Auckland, New Zealand

# Abstract

A rapid, cost-effective and easy method that allows on-site determination of the concentration of live and dead bacterial cells using a fibre-based spectroscopic device (the optrode system) is proposed and demonstrated. Identification of live and dead bacteria was achieved by using the commercially available dyes SYTO 9 and propidium iodide, and fluorescence spectra were measured by the optrode. Three spectral processing methods were evaluated for their effectiveness in predicting the original bacterial concentration in the samples: principal components regression (PCR), partial least squares regression (PLSR) and support vector regression (SVR). Without any sample pre-concentration, PCR achieved the most reliable results. It was able to quantify live bacteria from $10^8$ down to $10^{6.2}$ bacteria/mL and showed the potential to detect as low as $10^{5.7}$ bacteria/mL. Meanwhile, enumeration of dead bacteria using PCR was achieved between $10^8$ and $10^7$ bacteria/mL. The general procedures described in this article can be applied or modified for the enumeration of bacteria within populations stained with fluorescent dyes. The optrode is a promising device for the enumeration of live and dead bacterial populations particularly where rapid, on-site measurement and analysis is required.

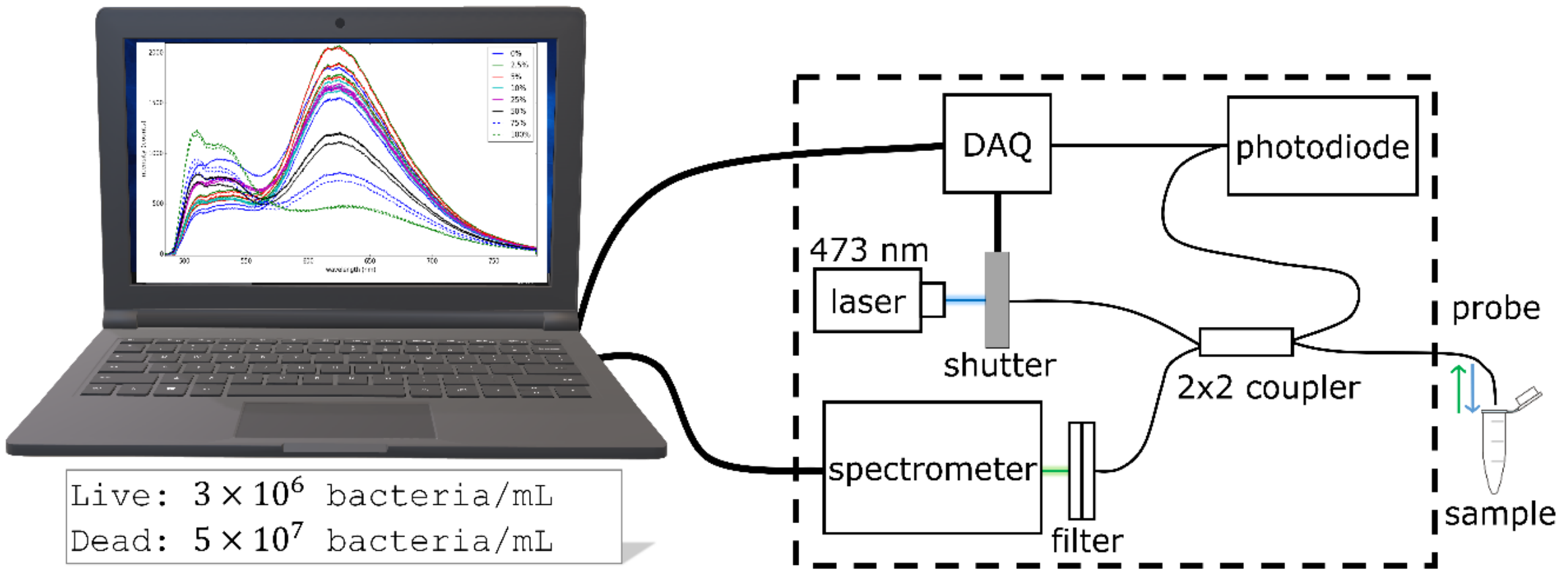

(Example output)

# Introduction

Quantifying microorganisms, and especially bacteria, is a vital in task in many fields of microbiology. Traditionally, bacterial viability is determined by the number of colonies (called colony forming units, or CFU) grown from a known volume on solid growth medium after a period of incubation. Inevitably, this method is labour intensive and involves a significant delay of 1 to 5 days. Furthermore, this method can only account for the cells that are culturable under the conditions of the experiment. Hence it cannot give an indication of the number of dead bacteria or the viable but non-culturable (VBNC) cells that retain their metabolic and cellular activity under stress [1]. In addition to detecting live bacteria, the enumeration and differentiation of dead bacteria is valuable or necessary in many applications. For example, in the evaluation of antimicrobial drugs [2,3], disinfection procedures [4], the viability of starter cultures [5], and monitoring of cell proliferation [3]. In all these applications, accurate and rapid information about the bacterial viability in the sample is desired.

Efficient, culture-independent detection of live and dead bacteria can be achieved using fluorescent dyes SYTO 9 and propidium iodide (PI) that differentially stain live and dead bacteria. Fluorescence detection is most commonly achieved by using microscopy, which allows direct investigation of individual cells. However, only a limited number of cells can be detected simultaneously, thus making the analysis of large sample volumes time consuming [6,7]. Fluorescence-based microplate readers offer more operational ease as the measurement of multiple samples can be automated and obtained in parallel [8]. The fluorescence intensity at discrete wavelengths are measured at the population level using optical filters or monochromats and extra calibration steps are required to obtain the sample concentration [9,10]. However, the accuracy of the calibration depends on the sensitivity of the plate reader and the quality of its optics, which both increase with cost [11]. Flow cytometry (FCM) allows study of cells at both the individual and population levels [12]. However, the application of FCM is restricted by its requirement of expensive and bulky equipment as well as trained technicians.

In this study, we describe a general method of using a fibre-based spectroscopic device called the optrode [13] to measure fluorescence from mixtures of live and dead *Escherichia coli* cells that are stained with SYTO 9 and PI. Compared to FCM and microplate methods, the optrode is cost-effective and easy to use while also having a more compact design. Selective sensitivity to enumerate specific bacterial populations can be achieved by using functionalised fibres or surfaces, as applied in other detection systems [14–16]. However, the standard optrode system does not require such sophisticated fabrication nor antibody activation. By dipping the fibre probe of the optrode directly into fluorescently tagged bacterial suspensions, the optrode accurately measures the emission signals at the cell population level.

The optrode allows versatile control of exposure times ranging from 8 ms to 10 s, suitable for the sensitive characterisation of various fluorophores. The optrode measures fluorescence spectra across the entire visible range, which is processed in this study to obtain information about the amount and state of bacteria in the samples. To demonstrate, the optrode was used to measure spectra from *E. coli* samples with concentration of $10^7$ or $10^8$ bacteria/mL, where the proportion of live:dead ranged from 0 to 100% live. Initially, the integrated fluorescence intensity of the SYTO 9 and PI emissions were directly used to model live and dead bacterial concentration, respectively. However, our results showed that the absolute intensities of dyes do not vary linearly across the range of bacterial concentrations investigated, and a more sophisticated processing method was required. Thus, the performance of three multivariate spectral processing methods were evaluated and compared: principal components regression (PCR), partial least squares regression (PLSR) and support vector regression (SVR).

## Results and Discussion

In this study, a portable and cost-effective fibre-based spectroscopic device (optrode) was used for the enumeration of fluorescently stained live and dead bacteria. Fluorescence spectra were collected from SYTO 9 and PI-stained *E. coli* samples where the proportion of live:dead ranged from 0 to 100% live. To model the concentration of live and dead bacteria in samples, the fluorescence spectra were used as predictors in three regression models: PCR, PLSR and SVR. We characterised the errors and investigated limits of detection for the described general optrode protocol in *E. coli* enumeration. This general protocol can be modified (e.g. by changing the volume or type of fluorescent dyes used) for the enumeration of a lower concentration range of *E. coli* or different types of bacteria.

### Fluorescence profile and interdependence of SYTO 9 and PI

The spectral training data ($N$ = 56 samples; $n$ = 159 measurements) were obtained from seven experiments where standard bacterial samples containing varying ratios of live:dead bacterial were prepared. The spectral profiles of stained bacteria exhibit peak maxima at *c.* 520 nm and 620 nm, corresponding to SYTO 9 and PI fluorescence, respectively. Figure 1a shows exemplar spectra to demonstrate the changes in peak intensity, position and shape when samples containing different proportions of live and dead bacteria are analysed for a total concentration of $10^8$ bacteria/mL. Initially, the integrated fluorescence intensity of SYTO 9 and PI were directly used to model the concentration of live and dead bacteria, respectively. However as shown in Figure 2, the dye intensities were variable and did not vary linearly with bacterial concentration. These results demonstrate that when used in combination, the intensity of SYTO 9 and PI varies with changes in the presence of live and dead bacteria in the sample, but these intensity variations are not directly proportional to the change in bacterial concentration. This complexity is expected due to the interactions of the nucleic acid dyes as they compete for the same target area and the possibility for Förster resonance energy transfer (FRET) to occur, which also has been documented in other studies [17,18]. Thus for this experiment, quantitative information about the presence of a bacterial subset cannot be obtained from direct correlation with the intensity of a fluorescence peak.

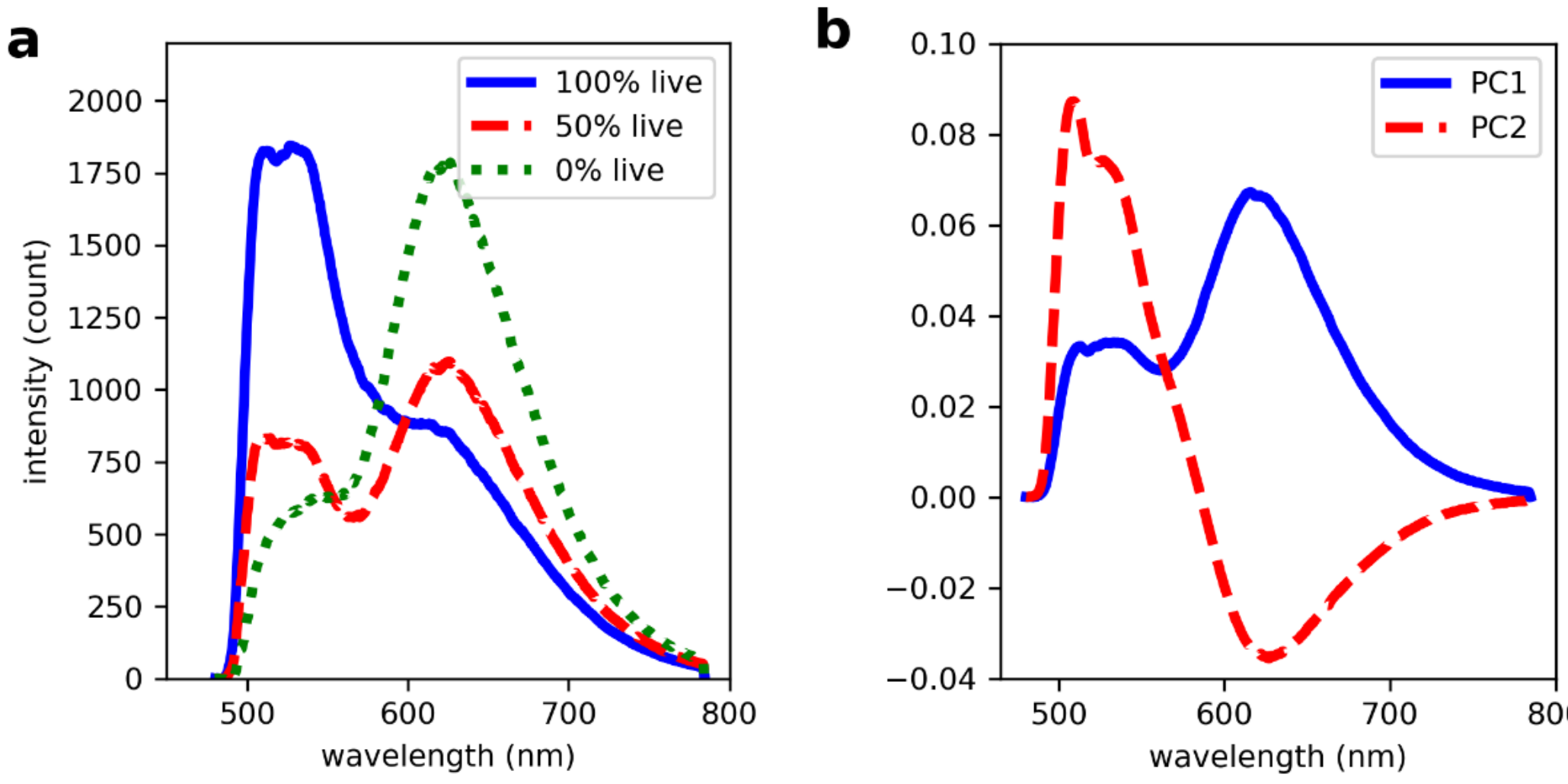

**Figure 1.** Measurements from the optrode system. (a) Exemplar spectra showing the difference in spectral profile measured from samples containing different ratios of live and dead bacteria. (a) The PCA signals of the spectral training dataset.

An alternative way to predict bacterial concentration from fluorescence spectra is to use information from the full spectrum instead of only the dye peak intensities. Analysis of the full spectrum will take into account features such as the relative intensity of both dyes, the shifts in wavelength and the changes in spectral shape, which provides information about the bacterial content of the samples. Numerous techniques including both traditional (e.g. PCR and PLSR) and machine learning approaches (e.g. SVR) allow multivariate inputs for the regression problem. To illustrate, the loadings plot in Figure 1b shows the principal component analysis (PCA) signal corresponding to each wavelength that contributed to the PCR analysis. PC1 and PC2 explained 65.6% and 34.2% of the variance in the spectral dataset, respectively. PC1 and PC2 both encompass the signals from SYTO 9 and PI, which is further evidence for the interdependence of these two dyes. The weights for the SYTO 9 and PI peaks are shaped differently between PC1 and PC2, which accounts for the spectral changes that occur as the proportion of live and dead bacteria varies. Compared to peak intensity alone, methods such as PCR, PLSR and SVR that use information from the entire spectral window allow better characterisation of the spectral changes relative to changes in bacterial content of the sample.

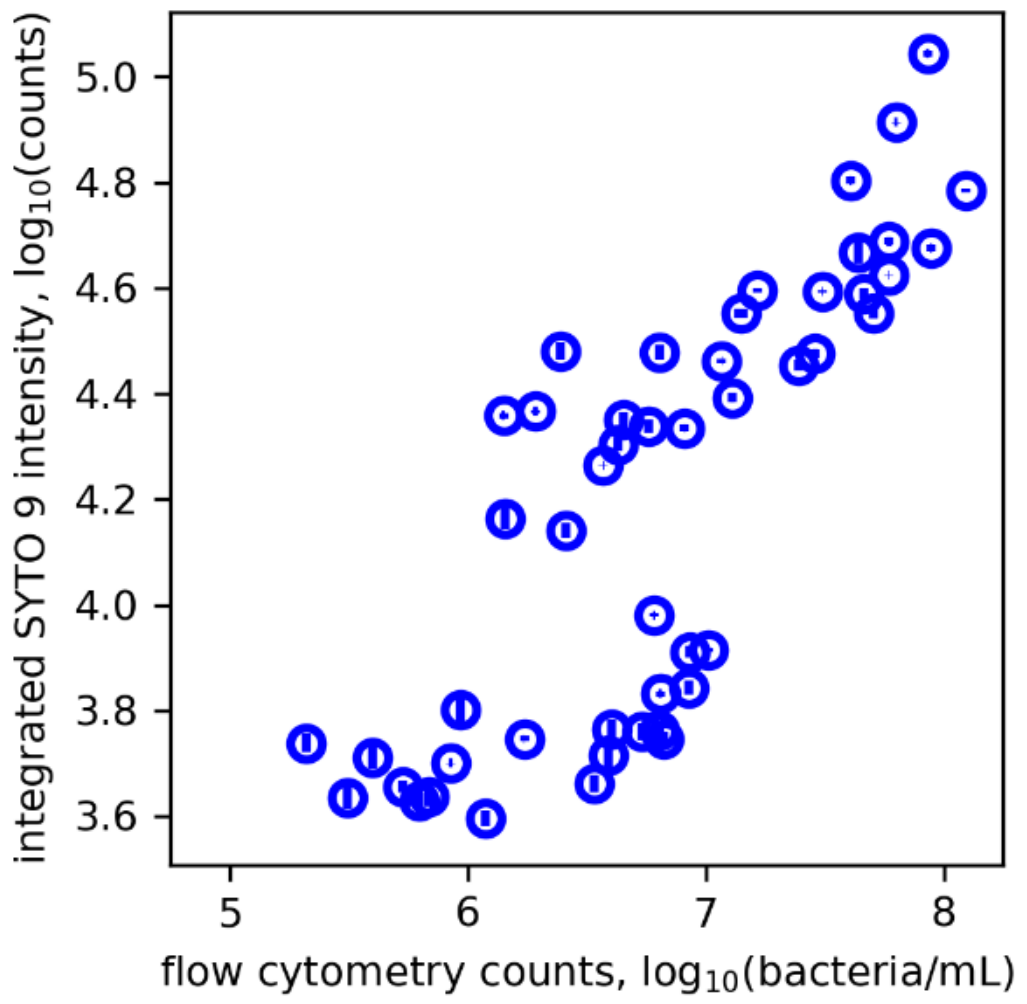

**Figure 2.** The integrated intensity of SYTO 9 obtained from optrode compared directly to the live bacterial concentration measured by FCM. The small vertical and horizontal error bars represent the standard error in replicate measurements. A log-log scale has been used to more clearly display the data.

## Comparison of the models: PCR, PLSR and SVR

The group K-fold cross validation (GKCV) analysis of PCR showed that the lowest root mean square error (RMSE) values for live and dead prediction models were 2 and 3 principal components (PCs), respectively. However, PCA decomposition of the spectral training dataset showed that the first two PCs accounted for more than 99.8% of the variance with PC3 only explaining 0.01% of the variance. As a result, the first two PCs were used in multiple linear regression to build the PCR models for predicting the concentrations of live and dead bacteria. On the other hand, two and four latent variables were included in the PLSR models of live and dead *E. coli*, respectively. The choice of latent variables for PLSR was based on the lowest values of RMSE from GKCV, which were 7.2 and 6.8 for the live and dead prediction models, respectively. For all three regression models, the predicted concentrations correlated linearly with the FCM-measured concentrations down to *c.* $10^{6.2}$ and *c.* $10^6$ bacteria/mL for live and dead bacteria, respectively. The $R^2$ and standard error of each regression model is summarised in Table 1, these values were calculated excluding data below the linear range.

| Model | $R^2$ | | Standard error | |
|---|---|---|---|---|
| | Live | Dead | Live | Dead |
| PCR | 0.77 | 0.88 | 0.25 | 0.17 |
| PLSR | 0.87 | 0.89 | 0.18 | 0.16 |
| SVR | 0.84 | 0.93 | 0.20 | 0.12 |

**Table 1.** Assessment of the PCR, PLSR and SVR models for the prediction of live and dead bacterial concentrations in training samples.

PLSR and SVR performed slightly better than PCR in modelling the live and dead bacteria in training samples, as they had higher $R^2$ and lower standard error values than PCR. However, the performance of PCR was better than both PLSR and SVR in predicting the concentration of test set samples, as shown in Table 2. PLSR and SVR are heavily dependent on identifying the precise patterns and relationships between the variations in spectral dataset and the expected bacterial concentration. However, the changes in spectral dataset become too subtle if the fluorescence signals are low, such as in the training samples with low concentrations or low percentages of the population of interest. As shown in the validation results, when there are spectral variations in the test set samples that are inconsistent with those in the training samples, PLSR and SVR predict bacterial concentration poorly and return invalid negative concentrations. In comparison, PCR is more robust against inconsistent spectral variations as shown by its performance in evaluating test set samples. Rather than focusing on the pattern between spectral predictors and target values, the predictive features used in PCR were chosen based on the amount of variance they explain in the spectral dataset. As PCR has less dependency on established patterns between predictors and targets, it demonstrated a superior ability to extrapolate the results for test set data.

## Validation of the models using test set samples

The regression models were validated and compared using external test set samples (*N* = 27 samples, *n* = 80 measurements) obtained from two blind experiments. The test set consisted of 24 samples within the concentration range of the training set (i.e. OD-estimated concentration of $10^7$ to $10^8$ bacteria/mL), and 3 extra samples with OD-estimated total concentration of $10^{6.5}$ bacteria/mL. All three regression models performed significantly better in predicting live than the dead bacterial concentration. PCR performed the best overall, in predicting the concentration of both live and dead bacteria in test set samples. In some instances, the models returned negative concentration values which were considered as invalid predictions. The results for the prediction of the test set samples are summarised in Table 2.

| Model | % of predictions within 2SE | | No. of invalid predictions | |
|---|---|---|---|---|
| | Live | Dead | Live | Dead |
| PCR | 92 | 56 | 2 | 0 |
| PLSR | 77 | 58 | 14 | 15 |
| SVR | 75 | 40 | 11 | 7 |

**Table 2.** Assessment of the PCR, PLSR and SVR models for the prediction of live and dead bacterial concentrations in test set samples. Invalid predictions refer to instances where the model returned negative concentration values.

Figure 3a shows the live bacterial concentration of test set samples analysed using PCR, which are mostly within the 95% confidence interval of the 1:1 line, as represented by the region of $\pm 2$ standard errors [19]. Enumeration of live bacteria in the 3 extra test set samples with OD-estimated total concentration of $10^{6.5}$ bacteria/mL was achieved. Furthermore, although the live bacteria in training samples modelled by PCR begin to flatten from *c.* $10^{6.2}$ bacteria/mL, test data with concentration down to $10^{5.7}$ bacteria/mL were predicted within 2 SE. Close inspection of the training data revealed that the subset which flattened from $10^{6.2}$ bacteria/mL contained low percentages of live bacteria. The FCM measurements on this subset of training samples showed that the proportion of live bacteria they contained were below 10% and 25% for the *c.* $10^8$ and $10^7$ bacteria/mL samples, respectively.

The PCR model predicted dead bacterial concentrations of test set samples are shown in Figure 3b. Although all three regression techniques modelled the dead bacteria in training samples down to *c.* $10^6$ bacteria/mL, none of the three models were able to predict the concentration in test set samples below *c.* $10^7$ bacteria/mL regardless of the proportion of live:dead bacteria that was contained.

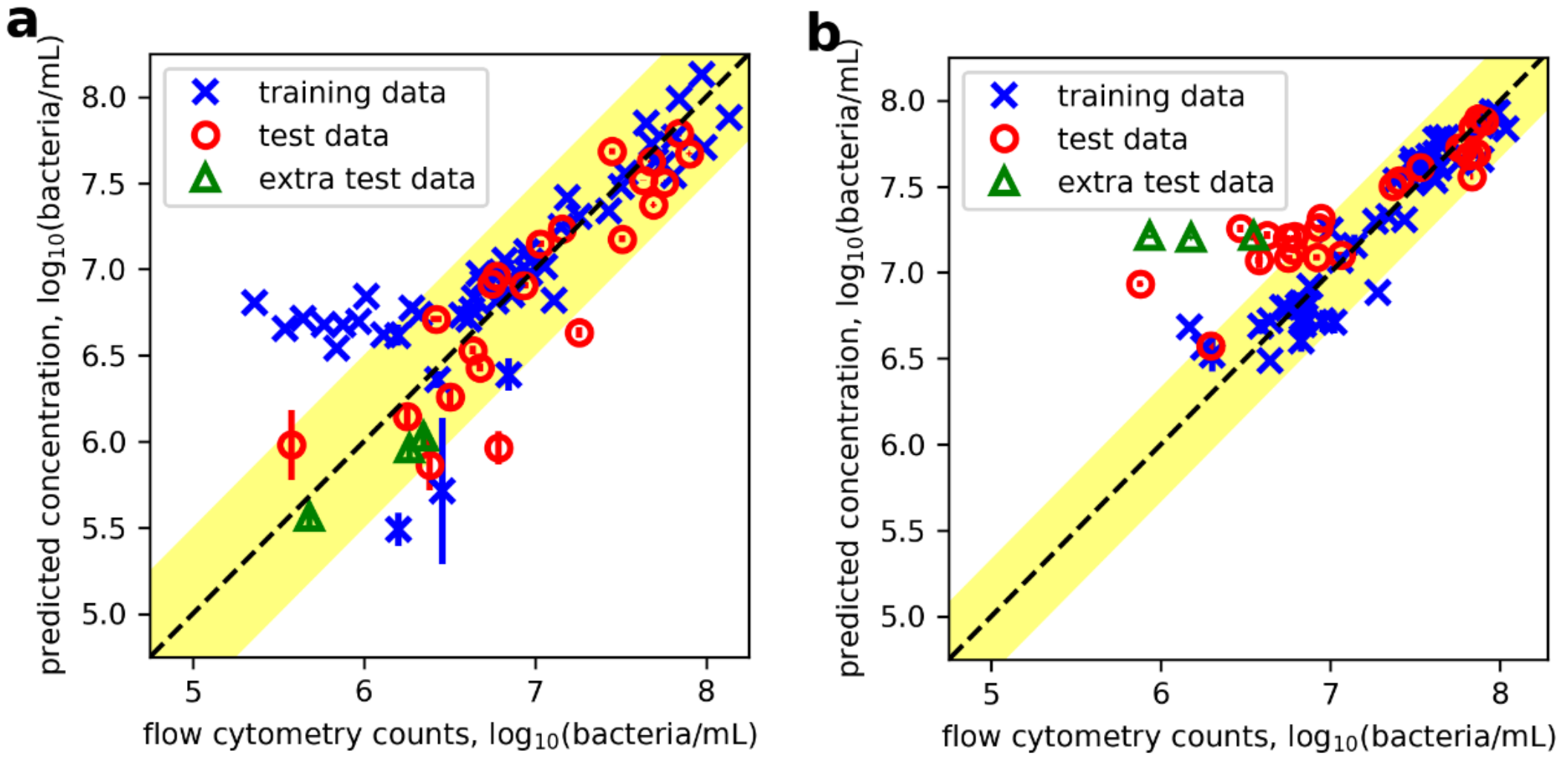

**Figure 3.** The log of the live (a) and dead (b) *E. coli* concentrations predicted using the PCR model (each model used 2 PCs) against the log of the *E. coli* concentrations measured by FCM. The spectral training data and test set data are represented by crosses and circles, respectively. Extra test samples containing lower total bacterial concentrations than the concentration range of the training set were also evaluated and represented in triangles. The dashed line marks the ideal 1:1 relationship between predicted concentration and that measured by FCM, and the shaded area represents the region of plus or minus two standard errors of the regression model. The vertical and horizontal error bars represent the standard error in replicate measurements. Samples with invalid predictions were excluded.

## Limit of detection (LOD) of the optrode method

The LOD of the PCR model for live and dead bacterial enumeration were found experimentally by examining their performance in modelling and predicting the training and test samples, respectively. PCR was able to model the live bacteria in training samples down to *c.* $10^{6.2}$ bacteria/mL. The samples below this threshold corresponded to the $10^7$ and $10^8$ bacteria/mL samples that contained 10% and 25% live, respectively. However, we observed that the test samples which contained higher %live bacteria were successfully predicted down to *c.*$10^{5.7}$ bacteria/mL. Thus, in samples with low proportions of live bacteria, the SYTO 9 signal can be overwhelmed by PI signal from the dead bacteria. This weak signal makes it difficult to obtain reliable measurements of the SYTO 9 signal, which contains the information about the presence of live bacteria. In addition, the amount of dyes added remained the same while the total bacterial concentration was lowered from $10^8$ to $10^7$ bacteria/mL. Thus, at lower bacterial concentrations there will be more dye available per cell which could potentially lead to more FRET. This will further decrease the SYTO 9 emission and increase the PI emission intensity in samples with lower concentrations, making it difficult to measure the live bacteria.

Overall, the regression models performed better at enumerating live bacteria than dead bacteria. One explanation for this is that in solutions of DNA, the increase in PI fluorescence upon binding is up to *c.* 9 times which is low compared to that of SYTO 9 which is more than 360 times [18]. The dead bacteria in training samples were linearly modelled by PCR down to *c.* $10^6$ bacteria/mL. However, the predictions of dead bacteria in test samples below *c.* $10^7$ bacteria/mL could not be achieved regardless of the %dead bacteria contained. When used in combination with SYTO 9, we observed that the peak intensity of PI in saline is comparable to that in dead bacterial samples at a concentration of $10^7$ bacteria/mL (Figure 4). Thus, to enumerate dead bacteria below *c.* $10^7$ bacteria/mL, the regression models would require information other than the intensity of PI, such as the changes of SYTO 9 intensity. However, these other spectral changes are not a direct result of the change in abundance of dead bacteria. For example, the decrease in SYTO 9 intensity would depend on the process of FRET, which depends on the abundance and proximity of PI, that then in turn depend on the abundance of dead bacteria and ease of access to its nucleic acids by PI. In addition, compared to SYTO 9 with its excitation maxima of 480 nm, PI has an excitation maxima of 540 nm and is less optimally excited using the 473 nm laser in the optrode [18]. With more optimal excitation of PI, the signals from the dead cells would appear stronger, more distinct and it may be possible to obtain more accurate predictions for dead bacterial concentration.

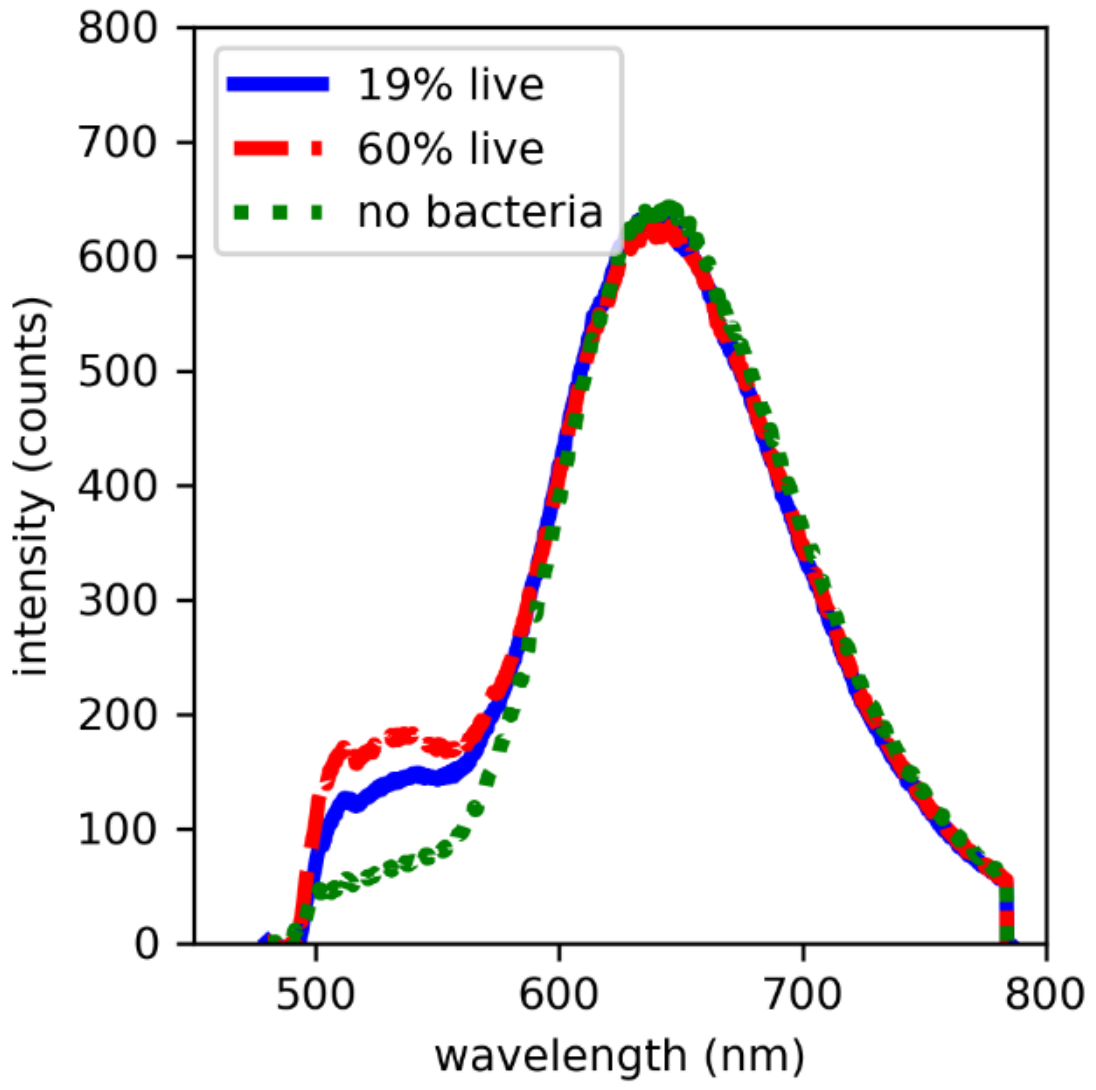

**Figure 4.** Exemplar spectra obtained from two fluorescently stained samples containing *c.* $10^7$ bacteria/mL, and one stained saline sample with no bacteria. There is an increase in SYTO 9 intensity in the samples containing bacteria. On the other hand, the intensity of PI in saline is comparable to that in samples containing different proportions of dead bacteria at concentration of $10^7$ bacteria/mL.

To investigate whether there were useful spectral patterns in samples that have concentration of live bacteria below the threshold of $10^{6.2}$ bacteria/mL, the spectral training data for these samples were analysed independently of the higher concentration samples. Out of PCR, PLSR and SVR, only PLSR showed the potential to model live bacterial concentration below *c.* $10^{6.2}$ bacteria/mL. However, evaluation using test set samples showed that the PLSR model built using low bacterial concentration data was unable to predict in this low concentration range. The fluorescence emission of the bacterial samples is affected by numerous factors including the ratio of dye molecules to nucleic acid (and hence the availability of unbound dye). As the stained bacterial samples were not washed to remove unbound dyes, the intensity change of spectral peaks in accordance to the changes in the low percentages of live or dead bacteria is limited. From the results we describe, the present method would favour situations where a controlled input concentration of bacteria is used to test the effectiveness of an antimicrobial intervention, e.g. a test of antimicrobial efficacy.

## Future improvements

In the future, adjustments can be made to improve the LOD that goes beyond the scope of the current study. For example, the volume of dyes used can be decreased to target the detection of low concentrations of bacteria, as this will help reduce background fluorescence. Washing the sample to remove unbound dye is an option, however this adds extra sample processing time and is difficult to achieve without losing bacteria in the process [20]. Suitability of dyes with higher quantum yields that selectively stain dead bacteria could also be explored [21]. In addition, it is worth investigating the relationship between changes in spectral shape and the percentage of live or dead bacteria present in a sample to provide information that can be used in combination with the predicted concentration to yield a more accurate and precise measurement of bacterial content.

Ultimately, the ability of the optrode technique to enumerate low concentrations of bacteria will be limited due to the dependence on the bacteria being in the collection volume of the fibre which is *c*. 0.028 $mm^2$ [22]. In this case, optrode faces the same difficulty as the microscope because to overcome the statistical counting error, a large number of measurements must be taken. To achieve this, the optrode could be programmed to sequentially record a series of measurements whereupon each series of spectral dataset can be analysed using a modified algorithm to account for the statistical fluctuations.

To conclude, a rapid, cost-effective and easy method that allows on-site determination of the abundance of live and dead bacteria is important in numerous fields including pharmacodynamic studies [2,3] and monitoring of cell proliferation [3]. In this study, we demonstrated that the fibre-based spectroscopic device (optrode) can be used to analyse a sample containing various ratios of live and dead *E. coli*, and obtain the concentration of each population. Of the three regression models investigated, PCR performed the best in predicting the live and dead bacterial concentrations in test set samples. The current optrode protocol with PCR is able to reliably enumerate live bacteria ranging from $10^8$ down to $10^{6.2}$ bacteria/mL, and there is potential to detect as low as $10^{5.7}$ bacteria/mL if there is a large proportion of live bacteria in the sample. On the other hand, enumeration of dead bacterial concentration can be achieved within the range of $10^8$ to $10^7$ bacteria/mL. By decreasing the volume of dyes used, the current optrode method can be adapted for the enumeration of lower bacterial concentrations.

The optrode is portable and requires little operator expertise which compares favourably with other forms of bacterial enumeration such as plate counting, fluorescence microscopy and flow cytometry. The optrode procedure takes about 20 min, with 15 min allotted for staining and the spectral measurements required less than 15 s. The method is potentially applicable to the enumeration of live and dead bacteria in a wide range of disciplines, particularly the medical and food industries.

# Material and methods

## Bacterial growth conditions.

*E. coli* strain ATCC 25922 (American Type Culture Collection, Virginia, USA) was incubated overnight in Difco tryptic soy broth (TSB; Fort Richard Laboratories, Auckland, New Zealand) then diluted 20 times and subcultured in fresh TSB for *c.* 1 h to yield a culture with an optical density (OD) at 600 nm, 1 cm path length of 0.6, equating to $4\times10^8$ bacteria/mL of exponentially growing cells. All broth cultures were grown at 37°C and aerated with orbital shaking at 200 rpm.

## Preparation of live:dead bacterial mixtures.

Bacterial suspensions were made using a modified protocol based on the instructions from the BacLight LIVE/DEAD Bacterial Viability and Counting Kit manual [23]. 10 mL of subcultured *E. coli* cells were harvested by centrifugation (4302 × *g*, 10 min, 21°C) followed by removal of supernatant and resuspension in 3 mL of saline (0.85% w/v). Subsequently, 1 mL of the harvested subculture was diluted in 9 mL of saline (live bacterial suspension) and another 1 mL diluted in 9 mL of 70% isopropyl alcohol (dead bacterial suspension). Each suspension was incubated for 1 h at 28°C and shaken at 200 rpm. Live and dead cells were harvested via three cycles of the washing process: centrifugation (4302 × *g*, 10 min, 21°C) followed by removal of supernatant and resuspension in 20 mL of saline. After the final wash, the cells were resuspended in saline to achieve a concentration of *c.* $10^8$ bacteria/mL; equivalent to diluting the sample to an OD of *c.* 0.2 at 600 nm. To make the standard bacterial samples, live and dead bacterial suspensions were combined in various volume ratios, giving *c.* 0, 2.5, 5, 10, 25, 50, 75, 100% live bacteria. The final bacterial suspensions contained varying live:dead ratios of either $10^8$ or $10^7$ bacteria/mL, with the total concentration estimated from OD measurements.

In addition to the set of standard bacterial samples for training the regression models, test set samples with OD-estimated total concentration of either $10^8$ or $10^7$ bacteria/mL were used as external validation for assessing the validity of the models. The live:dead ratios of the test samples did not coincide with that of the training set and to avoid bias, the preparation of the test set samples were blinded to the individual who measured and analysed them.

## Fluorescent dye staining and microsphere protocol.

BacLight LIVE/DEAD Bacterial Viability and Counting Kit (Invitrogen, Molecular Probes, Carlsbad, CA, USA; L34856) was used in our experiments. The kit contains a reference bead suspension and two nucleic dyes,

SYTO 9 and PI. SYTO 9 is membrane permeant while PI is membrane impermeant [23]. Saline was used for diluting the stock dyes to make working solutions of SYTO 9 and PI with concentration of 0.0334 mM and 0.4 mM, respectively. For each sample, 50 µL of the working solution of SYTO 9, 50 µL of the working solution of PI and 10 µL of homogenised reference beads were aliquoted into an empty microcentrifuge tube. Then, 900 µL of each bacterial sample was added to the dyes and beads, followed by gentle vortexing at 500 rpm in the dark for 15 min in room temperature. The final concentration of SYTO 9 and PI in each sample was 1.65 µM and 19.8 µM, respectively.

## Enumeration using the reference flow cytometry method.

All samples were evaluated using a LSR II Flow Cytometer (BD Biosciences, San Jose, CA, USA), with previously established protocols [20]. Briefly, excitation was achieved by a 488 nm laser with 20 mW power. SYTO 9 fluorescence was collected using a 505 nm longpass filter and a 530/30 nm bandpass filter. PI fluorescence was collected using a 685 nm longpass filter and bandpass filter with 695/40 transmission. Threshold was set to side scatter at 200. The flow rate was set to *c.* 6 µL/min and the duration of each measurement was 150 s. The number of microsphere beads added to each sample was used to calculate the absolute concentration of live and dead bacteria measured, via the bead-based FCM method [20,23,24].

This flow cytometry method was able to achieve reliable enumeration of live *E. coli* when its proportions ranged from 100% to 2.5% live; and reliable enumeration of dead *E. coli* in the concentration range of 100 % to *c.* 20% dead [20]. Thus to construct regression models, only the samples within the reliable enumeration range were used, i.e. the 2.5 to 100% and 0 to 75% live samples to model live and dead bacteria, respectively.

## Optrode system.

Bacterial fluorescence emission were recorded using a fibre-based spectroscopic system called the optrode [13]. Fluorescence excitation was achieved by a 473 nm solid state laser with *c.* 10 mW power. A data acquisition (DAQ) card synchronises the laser shutter with the spectrometer to minimise experimental variation from photobleaching. Using a 2x2 fibre coupler, laser light irradiates both the sample and a photodiode which monitors power fluctuations. A single probe made from multimode low OH silica fibre (diameter 200 µm, NA 0.22; Thorlabs Inc., Newton, NJ, USA) was used for excitation and fluorescence collection. The excitation line was removed by a 495 nm long-pass filter before reaching an Ocean Optics QE65000 CCD spectrometer which recorded the fluorescence spectra.

## Spectra acquisition and preprocessing.

The standard bacterial samples ($N$=56 samples and $n$=159 spectra, on average three per sample) and test set samples ($N$=27 samples and $n$=80, on average three per sample) were measured by the optrode with an integration time of 20 ms. The instrument dark noise was removed from each measurement, and the spectra were normalised to 10 mW laser power and 8 ms integration time. Then, the background spectrum acquired from saline was subtracted from each bacterial sample spectrum. Due to hasty cleaning of the probe in a few instances, one spectrum showed noticeably higher or lower intensity compared to the others that were recorded from the same sample, the abnormal spectrum was noted and excluded from the analysis. The remaining fluorescence spectra were mean-centered with respect to the average training spectra. These preprocessed fluorescence spectra were subsequently used in algorithms to correlate to bacterial concentration measured by FCM.

## Data analysis

Initially, the integrated intensity of SYTO 9 and PI were directly compared to the live and dead bacterial concentration, respectively. The regions of intensity integration corresponded to the fluorescence peak of the dyes bound to *E. coli*, which was between 509 – 529 nm for SYTO 9 and 609 – 629 nm for PI.

Subsequently, the performance of three regression techniques for predicting the concentration of live and dead bacteria were assessed and compared: PCR, PLSR and SVR. The three models were evaluated by external validation using test set samples. The $R^2$, standard error and the RMSE of the regression models were found. These algorithms were computed in Python codes using packages from NumPy [25], Matplotlib [26] and Scikit-learn [27]. Log-log plots (figure 3) were used to present the predicted versus measured *E. coli* concentrations clearly.

PCA was applied to reduce the multidimensional fluorescence spectral data to PCs, with the majority of the variance explained by the initial PCs. Multiple linear regression was performed using the initial PCs to build PCR models which correlates the spectral profiles to concentration of live and dead bacteria in the sample.

Similar to PCR, PLSR decomposes multidimensional input data into a smaller set of uncorrelated components called latent variables [28]. In contrast to PCA which decomposes the predictor data to obtain principal components that best explains variance in the predictor data itself, the first step of PLSR decomposes the predictor data to find latent variables that maximise covariance between predictor data and the response dataset (i.e. the expected values) [28]. This is followed by a regression step where a subset of latent variables are used to predict the response [28].

To determine the number of PCs and latent variables to use, the amount of variance explained and GKCV were used to evaluate the performance of PCA and PLSR models built with varying numbers of PCs and latent variables, respectively. The spectral training dataset was split into groups that corresponded to the seven experiments performed to collect the data. Seven iterations were performed for the GKCV and for each iteration, one group was held out as the internal test set while the remaining six were used as the training set. The number of PCs or latent variables that returned the lowest RMSE and appreciably increased the amount of variance explained were chosen.

The ε-SVR from Python's Scikit-learn library [27] was applied using the linear kernel. In SVR, the input spectral data are mapped onto a high-dimensional feature space through nonlinear mapping, and subsequently a linear regression model is constructed in this feature space [29]. Hyper parameters ε and C which defines the margin of tolerance and penalty factor, respectively, were determined using the grid search function with GKCV [27]. Grid search performed exhaustive search over various values of both parameters to find the best estimators that minimised the mean squared error of the predictions.

## Author information

**Corresponding author**

Fang Ou

Email: fou521@aucklanduni.ac.nz

**ORCID ID**

Fang Ou: 0000-0002-5037-0857

Cushla McGoverin: 0000-0003-3187-9743

Simon Swift: 0000-0001-7352-1112

Frédérique Vanholsbeeck: 0000-0001-9653-6907

**Notes**

The authors declare no competing financial interest.

## Acknowledgements

We are grateful to the New Zealand Ministry of Business, Innovation and Employment for funding the Food Safe; real time bacterial count (UOAX1411) research programme. This work is in partial fulfilment of Fang Ou's PhD thesis, who is grateful for the University of Auckland Doctoral Scholarship and the Todd Foundation Award for Excellence. The authors thank Stephen Edgar, Dr Julia Robertson, Zak Whiting and Janesha Perera for their laboratory support.